\documentclass[12pt]{spieman}  
\usepackage{amsmath,amsfonts,amssymb}
\usepackage{graphicx}
\usepackage{setspace}
\usepackage{tocloft}

\usepackage{pstricks}

 \hypersetup{breaklinks}

\title{General formalism for Fourier based Wave Front Sensing: application to the Pyramid Wave Front Sensors.}

\author[a,*]{Olivier Fauvarque}
\author[a]{Benoit Neichel}
\author[a,b]{Thierry Fusco}
\author[a,b]{Jean-Francois Sauvage}
\author[a]{Orion Girault}

\affil[a]{Aix Marseille Universit\'e, CNRS, LAM (Laboratoire d'Astrophysique de Marseille) UMR 7326, 13388, Marseille, France}
\affil[b]{ONERA--the French Aerospace Laboratory, F-92322 Ch$\hat{\text{a}}$tillon, France}

\cftpagenumbersoff{figure}
\cftpagenumbersoff{table} 

\begin{document} 
\maketitle

\begin{abstract}
In this article, we compare a set of Wave Front Sensors (WFS) based on Fourier filtering technique. In particular, this study explores the "class of pyramidal WFS" defined as the 4 faces pyramid WFS, all its recent variations and also some new WFSs as the 3 faces pyramid WFS.
Firstly, we describe such a sensors class thanks to the optical parameters of the Fourier filtering mask and the modulation parameters. 
Secondly, we use a unified formalism to create a set of performance criteria: size of the signal on the detector, efficiency of incoming flux, sensitivity, linear range and chromaticity.
Finally, we show the influence of the previous optical and modulation parameters on these performance criteria. This exhaustive study allows to know how to optimize the sensor regarding to performance specifications.
We show in particular that the number of faces has influence on the size of the signal but no influence on the sensitivity and linearity range. To modify these criteria, we show that the modulation radius and the apex angle are much more relevant. Moreover we observe that the time spent on edges or faces during a modulation cycle allows to adjust the trade-off between sensitivity and linearity range.
\end{abstract}


{\noindent \footnotesize\textbf{*}Olivier Fauvarque,  \linkable{olivier.fauvarque@lam.fr} }

\begin{spacing}{1}   

\section{Introduction}\label{sec:intro}

\begin{figure}[htbp]
\begin{center}
\includegraphics[width=16cm]{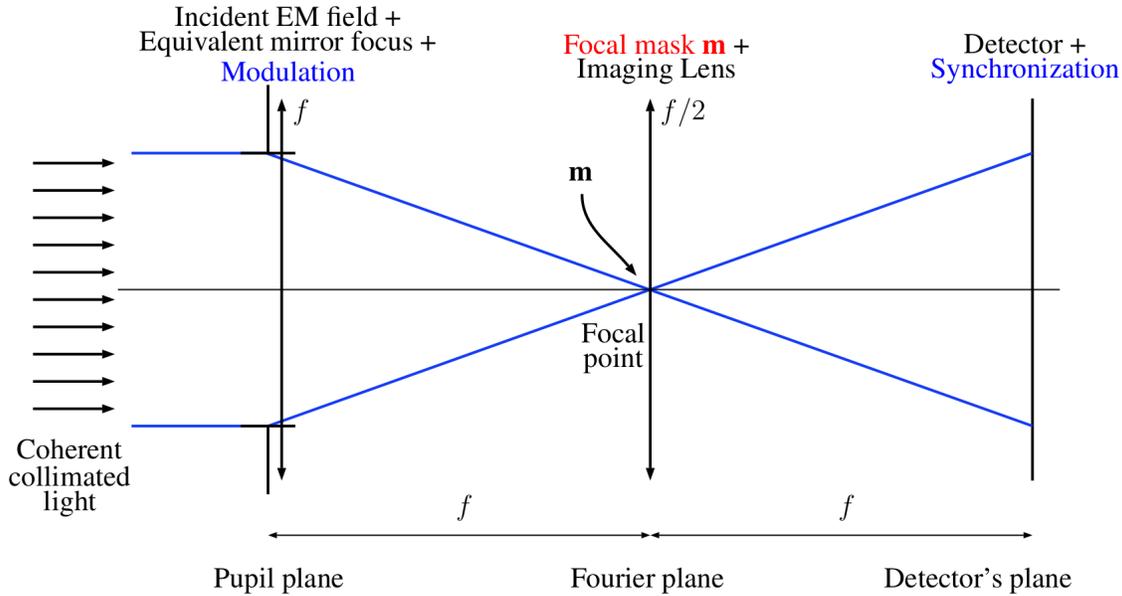}
\end{center}
\caption{Schematic view (in 1D) of a Fourier Filtering optical system. \label{ff_bench}}
\end{figure}

By placing amplitude or phase masks in a focal plane, it is possible to filter the light from a pupil plane to another. Those masks are able, in particular, to transform incoming phase fluctuations into intensity variations on a detector. Such optical designs (see figure \ref{ff_bench}) are thus particularly relevant in the context of Wave Front Sensing, especially for the Adaptive Optics (AO). Moreover, Fourier based Wave Front Sensors (WFS) have many advantages compared to others Wave Front Sensors as for example the Shack-Hartmann in terms of, for instance, noise propagation or sampling flexibility.

A general formalism about the Fourier based Wave Front Sensing has been recently developed in Fauvarque et al\cite{Fauvarque16}. Such a theoretical framework allows for instance to unify the Zernike Wave Front Sensor (WFS) introduced by Zernike\cite{zer1934} himself and the Ragazzoni's Pyramid WFS\cite{Rag96}. We choose in this article to explore, in the light of this formalism, the WFSs based on the Pyramid WFS principle, as for example the classical 4-faces modulated Pyramid, the 6- or 8-faces Pyramid introduced by Akondi et al.\cite{Akondi2014}, the Cone WFS (Vohnsen et al.\cite{Vohnsen2011}) or the Flattened Pyramid that we proposed in Fauvarque et al.\cite{Fauvarque2015}. The choice of studying this class of Fourier-based WFSs is due to the fact that the PWFS recently shows its great efficiency on sky on the Large Binocular Telescope (Esposito et al.\cite{LBT}), the Magellan Telescope (Close et al.\cite{MagAO}) and the Subaru Telescope (Jovanovic et al.\cite{Subaru}). It subsequently seems to be the most credible candidate for the next generations of Adaptive Optics, especially for the European Extremely Large Telescope. As a consequence, serious optimization works will have to be led in order to know which designs will be the most appropriate regarding to the Wave Front Sensing contexts. Finally, we mention some of the most enlightening theoretical works about this PWFS: Ragazzoni et al.\cite{ragazzoni1999sensitivity}, Esposito et al.\cite{Esposito1999}, Verinaud\cite{verinaud2004nature}, and Guyon\cite{guyon2005limits} which will serve as reference results along the article.

In the first part of this paper, we describe the "Class of Pyramidal WFSs" which contains the modulated Pyramid WFS and all its variation when considering optical parameters of the mask and modulation settings as free parameters. The second part will define the unified performance criteria which will be used in the third part to compare all these Wave Front Sensors. In terms of optimization approach, the first part corresponds to the input parameters while the second part describes the output specifications. A final part will summarize the influence of each parameters on each performance criteria in order to create the best WFS regarding to an AO context.

We recall here the general framework of the Fourier based Wave Front Sensing. The optical design is described in figure \ref{ff_bench}.
\vspace{0.5cm}
Such a device is a typical Fourier filtering system. The incoming perturbed electro-magnetic field $\psi_p$ is written:
\begin{equation}
\psi_p(\phi,n) = \sqrt{n}~\mathbb{I}_P~exp \left( \imath\phi \right)
\end{equation}
where $n$ is the spatial averaged flux, $\phi$ the perturbed phase at the analysis wavelength $\lambda_0$, and $\mathbb{I}_P$ is the indicative function of the entrance pupil.
The Fourier mask which takes place on the focal point is considered as a pure transparent mask. Its transparency function may be written:
\begin{equation}
m = exp\left(\frac{2\imath\pi}{\lambda_0}  OS\right)
\end{equation}
where $OS$ is the Optical shape of the mask.
Via the Fraunhofer optical formalism, it is possible to write the intensity on the detector:
\begin{equation}
I(\phi,n) =|\psi_p(\phi,n) \star \mathcal{F}[m]|^2
\end{equation}
where $\mathcal{F}[m]$ is the 2D Fourier transform of the mask.
%
The phase seen by the WFS is the sum of the turbulent phase induced by the atmosphere and the static aberrations of the wave front sensing path. 
Mathematically, we can split the incoming phase into two terms:
\begin{equation}
\phi = \phi_r + \phi_t
\end{equation}
where $\phi_t$ is the turbulent phase and $\phi_r$ the static reference phase. $\phi_r$ may also be seen as the operating point of the WFS. 
Thanks to a Taylor's development of the phase around the reference phase $\phi_r$:
it is possible to get a phase power series of the intensity on the detector:
\begin{equation}
I(\phi,n)=n\Big(I_c+I_l(\phi_t)+I_q(\phi_t)+...\Big)
\end{equation}
The first term $I_c$ is the \emph{constant} intensity, it corresponds to the intensity on the detector when the phase equals to the reference phase, i.e., when the turbulent phase is null. The second term is the \emph{phase-linear} term. It corresponds to the perfectly linear dependence of the intensity regarding  to the turbulent phase around the reference phase. The third term $I_q$ is the quadratic intensity, it corresponds to the first non-linear dependence of the intensity. The next terms are obviously non linear contributions as well. These equations may be easily generalized when a modulation device is working. The exact expressions of these intensities are given in Fauvarque et al.\cite{Fauvarque16}

Since the signal on the detector is not linear with the phase, we have to process it. The easiest way to create a phase-linear quantity from $I(\phi,n)$  is to calculate the meta-intensity, called $mI$, via the following equation:
\begin{equation}
mI(\phi_t)= \frac{I(\phi_r+\phi_t,n)-I(\phi_r,n)}{n}\label{MI}
\end{equation}
The normalization by the factor $n$ allows to make $mI$ independent from the incoming flux. Retrieving $I(\phi_r,n)$ corresponds to a \emph{tare} operation. In practice, this \emph{return-to-reference} operation is done via a calibration path. This ensures that the meta-intensity equals to zero when there is no turbulent incoming phase. With such a definition, the meta-intensity $mI$ equals to the  
\emph{linear} intensity $I_l$ in the small phases approximation regime.

\section{The Pyramid Wave Front Sensors Class}\label{free_para}

We introduce in this section the Pyramid Wave Front Sensors class. This sensors class corresponds to the generalization of the classical Pyramid WFS introduced by Ragazzonni\cite{Rag96} when you consider the number of faces and the apex angle of the mask and the modulation settings as free parameters. 

\subsection{Number of faces}

\begin{figure}[htbp]
\includegraphics[width=16cm]{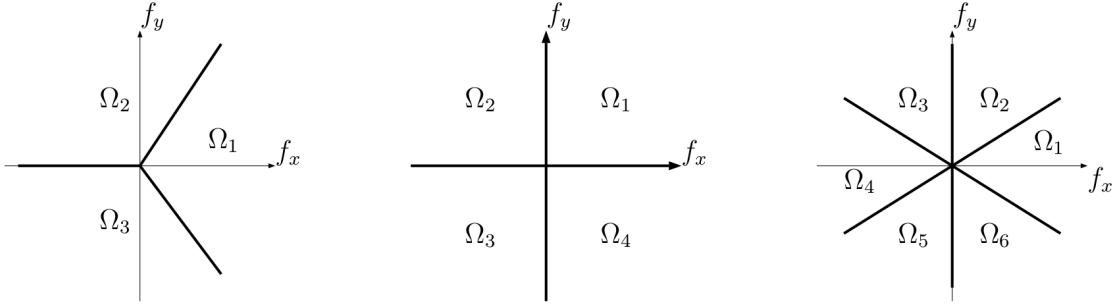}
\caption{Different types of Fourier plane tessellation for the Pyramid WFS class: 3, 4 and 6 faces.\label{Tesss}}
\end{figure}

\begin{figure}[htb]
\centering
\includegraphics[width=14cm]{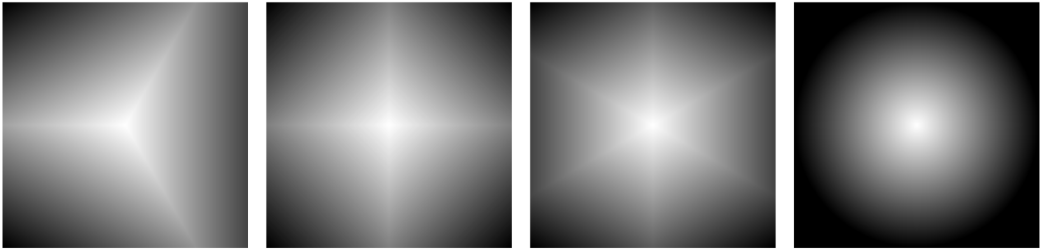}

\caption{Optical shape of the studied masks. From left to right: 3, 4, 6 and $\infty$ faces. The 4 faces pyramid corresponds to the classical optical configuration.  The last mask is equivalent to the axicon. \label{stu_mask}}
\end{figure}

Since the focal plane corresponds to a Fourier plane, the number of faces will set the tessellation of the spatial frequencies plane. In the case of the PWFS class, we will only consider tessellations which are centered on the focal point and provide equal partitions of the energy. Moreover, since we want to describe the Fourier plane we need at least two vectors which are not collinear, i.e. three points which are not aligned. As a consequence the minimal tessellation corresponds to a 3-faces pyramid. The opposite case corresponds to a conic pyramid, i.e. a pyramid with an infinite number of faces. This design has the advantage to not have privileged axis. Figure \ref{Tesss} show the tessellation of the focal plane associated of the 3, 4 and 6 faces pyramids (see their optical shapes on the three left inserts of figure \ref{stu_mask}). 

\subsection{Apex angle}
If the number of faces allows to understand how the spatial frequencies are split, we need to explore the apex angle parameter to know how these frequencies are rejected. Indeed, the slopes of each faces (which are local tip/tilt angles) will set the place where the pupil images will be. If the apex angle is large enough, the spatial frequencies filtered by the tessellation will be completely split. However, if the angle is smaller, the pupil images will overlap and create interferences. Such a design have been studied in letter\cite{Fauvarque2015} for a number of faces equals to 4 and an overlap rate of the pupil of 90\%. Another sets of parameters will be tested in this article.

\subsection{Modulation path}

\begin{figure}[htb]
\centering
\includegraphics[width=13cm]{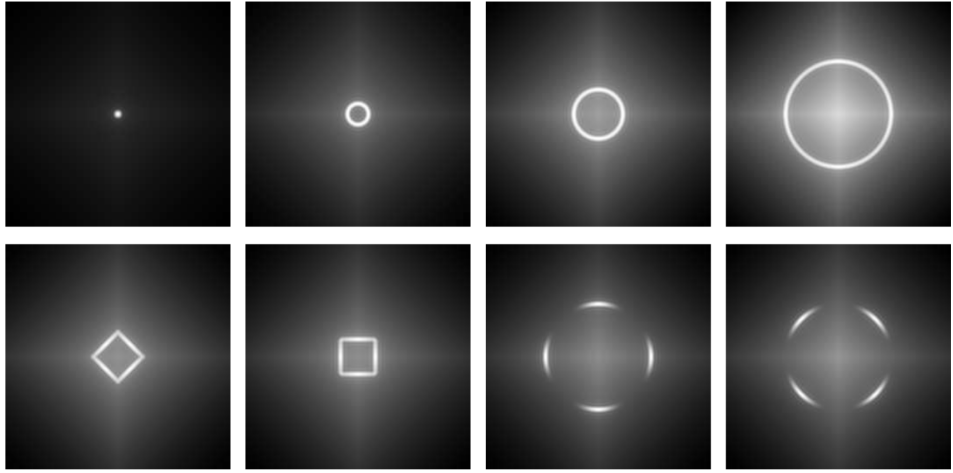}~

\caption{Some examples of modulation paths generated by changing the modulation radius and the weighting function.\label{modd}}
\end{figure}

In addition to the number of faces and the apex angle, we also consider the modulation path as a free parameter. For a tip/tilt modulation, the general expression of the modulated phase $\phi_m$ is:
\begin{equation}
\phi_m(s) = r_m(s) \left[\cos(2\pi s) Z_1^{-1} + \sin(2\pi s) Z_1^{1}\right]w(s)
\end{equation}
$Z_1^{1}$ and $Z_1^{-1}$ are the Tip and Tilt Zernike polynomials.
$s$ is the temporal variable normalized with respects to the duration of a modulation cycle. $r_m(s)$ codes the modulation radius. $w(s)$ is the weighting function which allows to spend more time on the faces or on the edges. The intensity associated to the modulation is:
\begin{equation}
I(\phi,n) =\int |\psi_p(\phi + \phi_m(s),n) \star \mathcal{F}[m]|^2 w(s) ds
\end{equation}
By changing $r_m(s)$ and $w(s)$ it is possible to generate an infinite number of tip/tilt modulation. A constant modulation radius define a circular modulation but a squared path may also be coded. A weighting function which equals to 1 all along the modulation path is called \emph{uniform modulation}. Such a case is usually used in the existent modulation devices but, obviously, such a constrain may be released in order to change the WFS properties. Some examples of modulation paths are given on figure \ref{modd}.

\subsection{Interest area of the detector}

In this article, we only take into account photons which are in the geometrical footprint of the entrance pupil images. It means that we do not consider the diffracted light.

\begin{figure}[htbp]
\centering
\includegraphics[width=11cm]{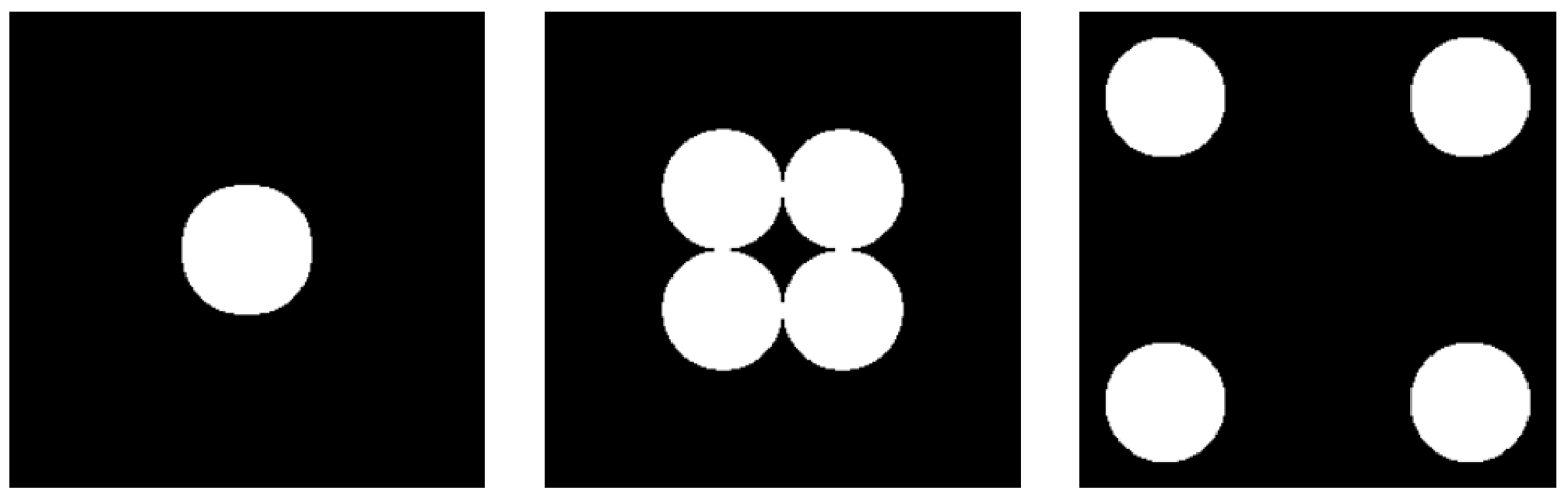}
\caption{Interest Area for a circular pupil for the 4-faces PWFSs. From left to right: $\frac{2f\theta}{D}$=0.1, 1 and 3. $D$ is the diameter of the circular pupil and $\theta$ codes the apex angle. \label{Pup_ecart}}

\medskip

\includegraphics[width=15cm]{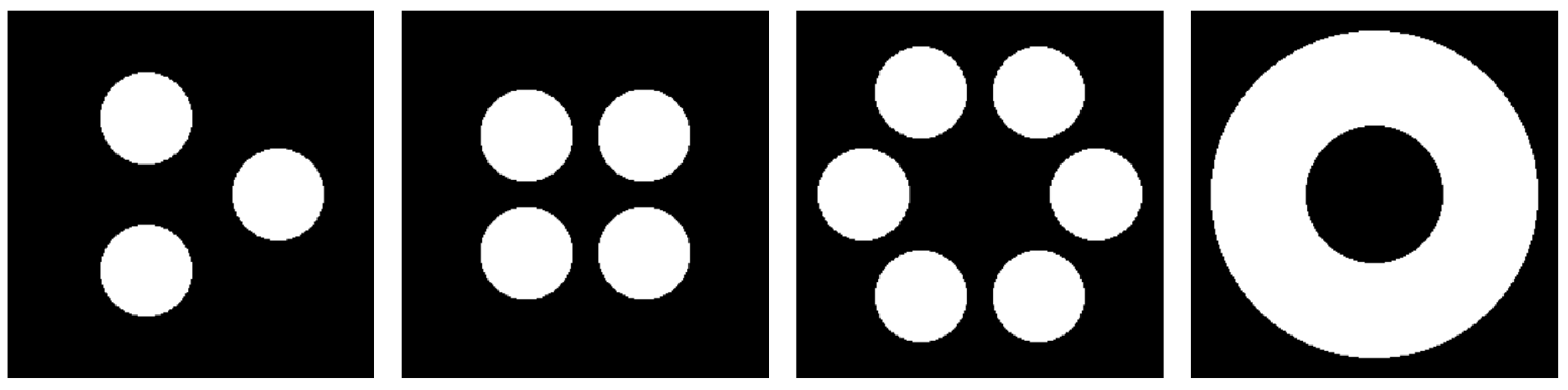}\\
\caption{Interest Area for a circular pupil for the PWFS class. From left to right: 3, 4, 6 faces and the cone. Apex angle $\theta$ equals to $1.5\frac{D}{2f}$.  \label{Pup_ecart2}}
\end{figure}

The geometric images of the pupil are easy to obtain: they are totally characterized by the tip/tilt angles of each face $\Omega_i$. If we call $(\alpha_i,\beta_i)$ such angles associated, the pupil image of the tessellation $\Omega_i$ will take place around the detector's point $(f\alpha_i,f\beta_i)$ where $f$ is the focal of the imaging lens. The interest pixels, i.e. the geometrical footprints of the entrance pupil, are thus the ones which are within a pupil radius of the points $(f\alpha_i,f\beta_i)$.  The indicative function of the interest area will be called $\mathbb{I}_{IA}$.
We show on figures \ref{Pup_ecart} and \ref{Pup_ecart2} the interest area for different optical parameters. 

\section{Performance criteria}

We introduce in this section the performance criteria which will be used in section \ref{comparison} to compare the wave front sensors of the PWFSs class when you change the parameters of section \ref{free_para}. 

\subsection{Surface ratio}

\begin{figure}[htbp]
\centering
\begin{minipage}[c]{0.5\linewidth}
\centering
\rotatebox{90}{\hspace{2cm}Surface ratio}
\includegraphics[trim = 2.2cm 1.5cm 1.75cm 2cm, clip,width=7.5cm]{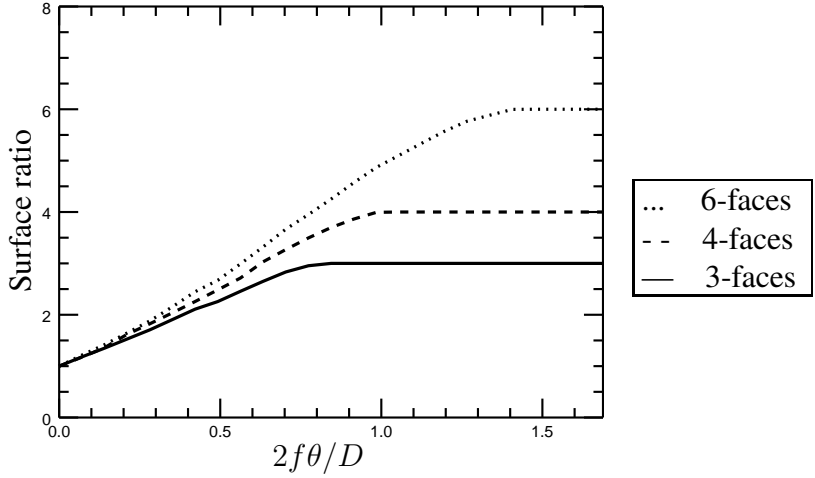}\\
\vspace{-0.2cm}
~~~$2f\theta/D$
\end{minipage}~
\fbox{\begin{minipage}[c]{0.13\linewidth}
... ~~ 6-faces

- - ~~ 4-faces

--- ~~ 3-faces
\end{minipage}}
\vspace{0.2cm}
\caption{Surface ratio for the Pyramid WFSs class with an increasing apex angle $\theta$.\label{sigma}}
\end{figure}

The first one is about the number of detector's pixels used to code phase information. Ideally, this number equals to the number of pixels of the entrance pupil. It means that each pixel codes one phase mode. Technologically, it also means that the detector optimizes each of its pixel. The mathematical quantity which illustrates this criterion corresponds to the ratio between the surface of the interest area and the surface of the entrance pupil: 
\begin{equation}
\sigma = \frac{||\mathbb{I}_{IA}||_1}{||\mathbb{I}_{P} ||_1}
\end{equation} 
This quantity will be called the \emph{surface ratio}. In the present context it, at least, equals to 1. It's the case of the completely flat PWFS which unfortunately generates a useless WFS. The classical Pyramid WFS has a surface ratio of 4. The Flattened 4-faces pyramid has a transitional surface ratio going from 1 (when the pyramid is completely flat) to 4 when the angle apex increases. Figure \ref{sigma} shows the \emph{surface ratio} with respects to the apex angle for the 3, 4 and 6-faces pyramids.

\subsection{Efficiency of the incoming flux}

The second criterion quantifies the percentage of the incoming flux which is effectively used to code phase information. Since the reference sources are quite faint in the context of astronomy, such a percentage is a strong indicator of the WFS performance. Even tough the number of photons in the interest area is dependent on the incoming phase, it appears that this dependence is weak (at least in the small phases approximation). That's why, we choose to count the flux in the interest area when the phase equals to the reference phase to estimate this efficiency rate. Mathematically, this rate called $\eta$ is:
\begin{equation*}
\eta = \frac{||\mathbb{I}_{IA}.I(\phi_r,n)||_1}{n}
\end{equation*}
This quantity goes from 0 to 1. A perfect wave front sensor has obviously a photon efficiency rate of 1 whereas it would equal to 0 for a perfect coronagraph. 
\begin{figure}[htbp]
\begin{minipage}[c]{0.48\linewidth}
\centering
\rotatebox{90}{\hspace{1.2cm}Photon efficiency rate}
\includegraphics[trim = 1.8cm 1.5cm 1.75cm 2cm, clip,width=7cm]{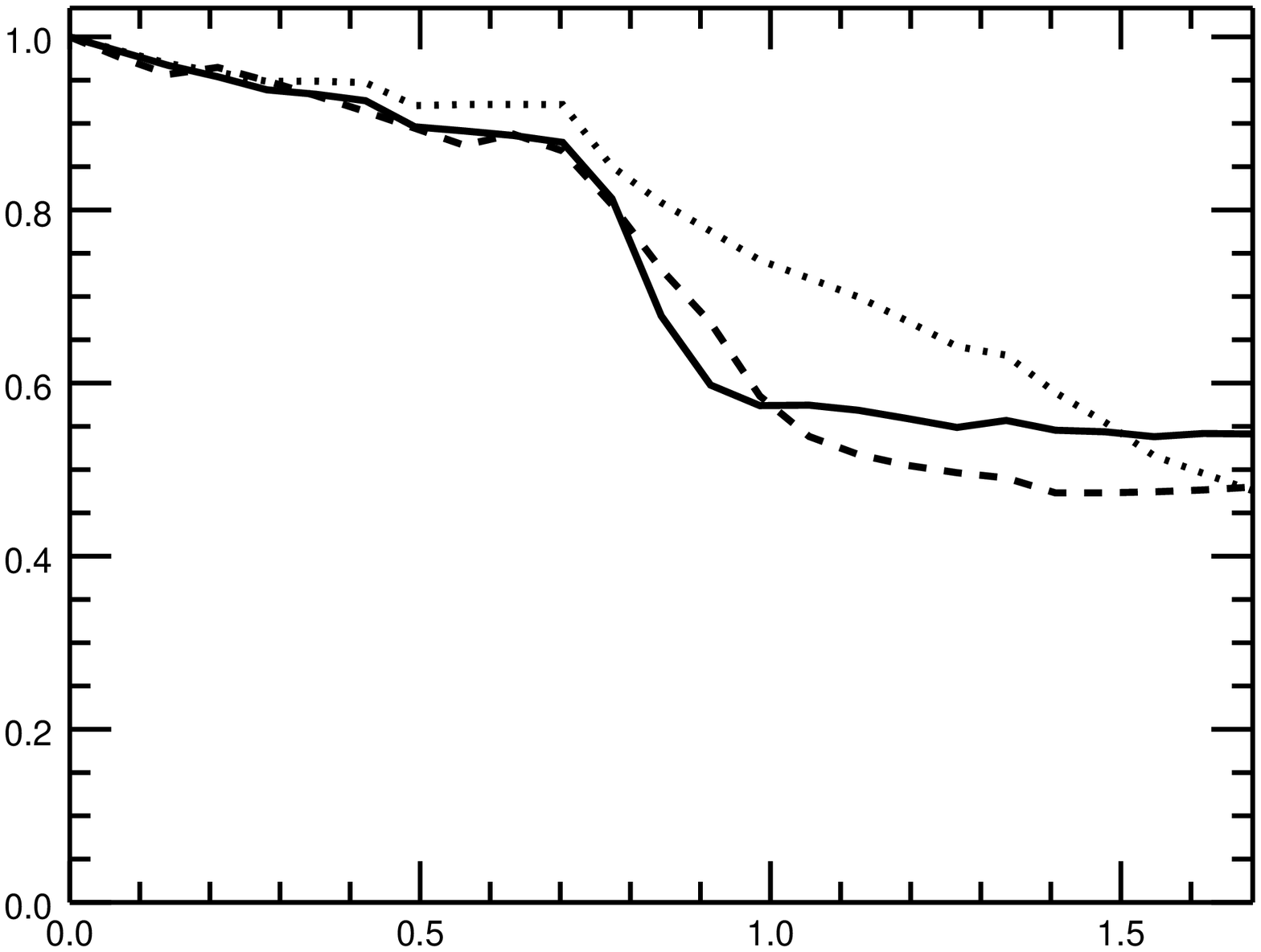}\\
\vspace{-0.2cm}
~~~$2f\theta/D$
\caption{Photon efficiency rate for the Pyramid WFSs class  \textbf{without modulation} when the apex angle $\theta$ is increasing. \label{eta_m0}}
\end{minipage}~~
\hspace{-6cm}\fbox{\begin{minipage}[c]{0.10\linewidth}
\begin{scriptsize}
... ~~ 6-faces

- - ~~ 4-faces

\vspace{-0.2cm}

--- ~~ 3-faces
\end{scriptsize} 
\end{minipage}}~~\hspace{4cm}
\begin{minipage}[c]{0.48\linewidth}
\centering
\rotatebox{90}{\hspace{1.2cm}Photon efficiency rate}
\includegraphics[trim = 1.8cm 1.5cm 1.75cm 2cm, clip,width=7cm]{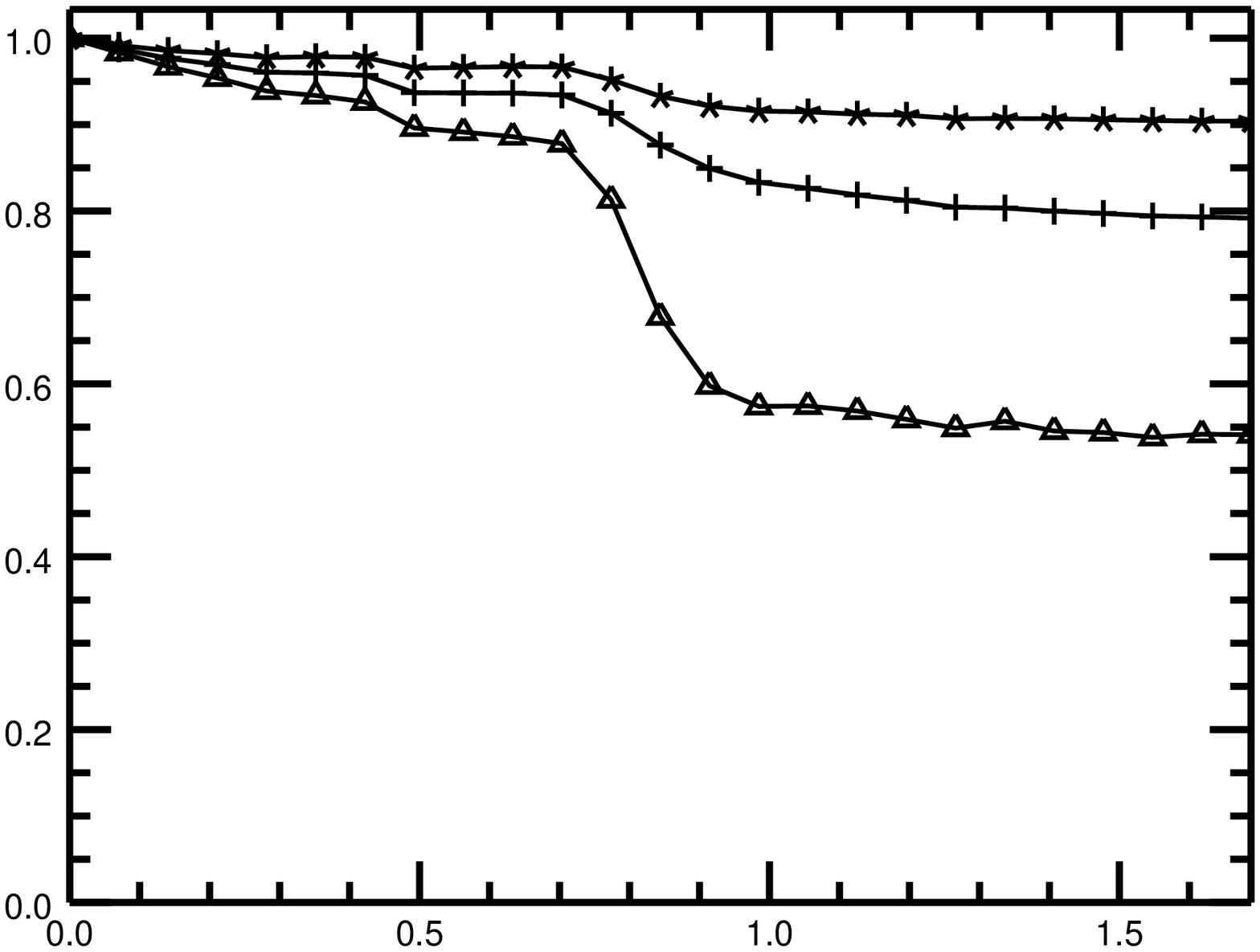}\\
\vspace{-0.2cm}
~~~$2f\theta/D$
\caption{Influence of the modulation radius on the photon efficiency rate for the \textbf{3}-faces Pyramid WFSs class with an increasing apex angle $\theta$.    \label{eta_m}}
\end{minipage}
\hspace{-6cm}\fbox{\begin{minipage}[c]{0.13\linewidth}
\begin{scriptsize}
$*$~~~~$r_m=3\lambda/D$

+  ~~$r_m=1\lambda/D$
\vspace{-0.1cm}
$\triangle$ ~~$r_m=0\lambda/D$
\end{scriptsize} 
\end{minipage}}
\end{figure}
Figure \ref{eta_m0} shows that the Flattened Pyramids (3, 4 or 6 faces) have almost perfect photon efficiency rates:  higher than 90\% for $2f\theta/D \in [0,0.5]$ while $\eta$ drastically decreases around 50\%  whatever the number of faces equals to when the pupil images do not overlap anymore. Figure \ref{eta_m0} shows that modulation allows to retrieve a great part of the lost photons, especially when the pupil images are completely separated.

\subsection{Chromaticity}
We use the definition of an \textbf{achromatic sensor} introduced in article\cite{Fauvarque16}. The influence of an incoming polychromatic light on the output meta-intensities for such a sensor is only in terms of gain and not in terms of structure. In other words, there is no cross talk between the turbulent phase modes due to the incoming polychromatic light. In order to know if a sensor is achromatic, we use the "substitution test" developed in Fauvarque et al.\cite{Fauvarque16}. Such a test consists in checking the capability to make the transparency function of the mask independent of the wavelength $\lambda$ via the following substitution:
\begin{equation}
(u,v) = \left(\frac{f_x}{\lambda},\frac{f_x}{\lambda}\right) \label{sub}
\end{equation}
For the 4 faces pyramid the transparency function is:
\begin{eqnarray}
m_4(f_x,f_y,\lambda)=&\exp\left(\frac{2\imath\pi}{\lambda}\theta(|f_x|+|f_y|)\right)\\
m_4(u,v)=&\exp\left(2\imath\pi\theta(|u|+|v|)\right)
\end{eqnarray}
Since the wavelength dependence disappears thanks to substitution \ref{sub}, the 4-faces PWFS is thus an achromatic sensor. Moreover, there is no constraint on the apex angle $\theta$ concerning the wavelength, as a consequence, the Pyramid WFS is achromatic whatever the apex angle is. One can generalize this result for any number of faces. Unfortunately, Pyramid WFSs lose this achromatic behavior as soon as the reference phase is not null anymore or when the modulation is used.

\subsection{Manufacturing}
\begin{figure}[htpb]
\centering
\includegraphics[width=10cm]{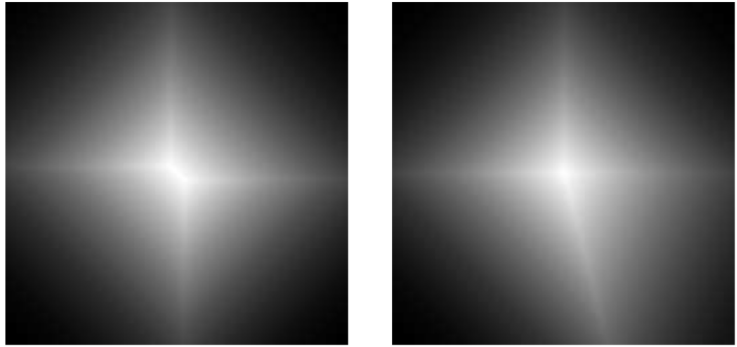}
\caption{Manufactured 4-faces pyramids with classical defaults: (Left) Roof-top shape. (Right) Asymmetric angles. \label{manu_pyr}}
\end{figure}

A relevant criterion is about the ease of manufacturing the pyramidal mask. We know that imperfections are recurring for the 4-faces Pyramid as for example the "roof-top shape" (see left insert of figure \ref{manu_pyr}) instead of a perfect pyramidal shape. These difficulties are due to the fact that a point is defined as the intersection of three planes and not 4. In others words, a 3-faces pyramid is much easier to manufacture than a 4-faces. Moreover, it might be technologically difficult to equally separate the faces (see right insert of figure \ref{manu_pyr}) or to create very sharp edges, that's why using a Cone as filtering mask could be particularly convenient.

\subsection{Sensitivity, Linearity range and SD factor}
Finally, we are interested in the crucial criteria which are sensitivity and the linearity range.

The sensitivity associated to a normalized\footnote{regarding to the RMS norm.} turbulent mode~$\phi_t$, that we call $s(\phi_t)$, is calculated directly from the \emph{linear} intensity of the phase power series of the intensity since this quantity corresponds to the perfectly linear behavior of a WFS:  
\begin{equation}
s(\phi_t) = ||I_l(\phi_t)||_2 \label{sensii}
\end{equation}
The linear range, that we call $d(\phi_t)$, is calculated via the first non-linear term, i.e. the \emph{quadratic} intensity $I_q$. The higher this term is, the smaller the linearity range will be. 
\begin{equation}
d(\phi_t) = \frac{1}{||I_q(\phi_t)||_2} \label{dynaa}
\end{equation}
We also consider the product of these two antagonist performance criteria that we call the \textbf{SD factor} (for Sensitivity versus Dynamic) since it quantifies the trade-off between them.
The SD factor is particularly relevant to know, for instance, what is the best way to improve the linearity range without too much decreasing the sensitivity.

\section{Comparison of the Pyramid Wave Front Sensors}\label{comparison}
\begin{figure}[htpb]
\centering
\includegraphics[width=14cm]{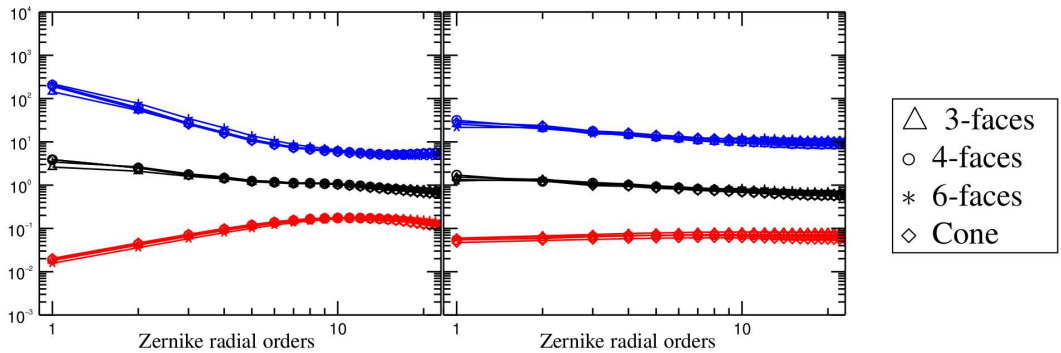}

\caption{Influence of the \textbf{number of faces} on the \textcolor{red}{\textbf{Sensitvity}}, the \textcolor{blue}{\textbf{Linearity range}} and the \textbf{SD factor} with respect to the spatial frequencies in terms of Zernike Radial Orders.  
Left insert: small apex angle ($\theta=0.1\frac{D}{2f}$). Right insert: large apex angle ($\theta=2\frac{D}{2f}$). \label{1}}
\end{figure}

\begin{figure}[htpb]
\centering
\includegraphics[width=16cm]{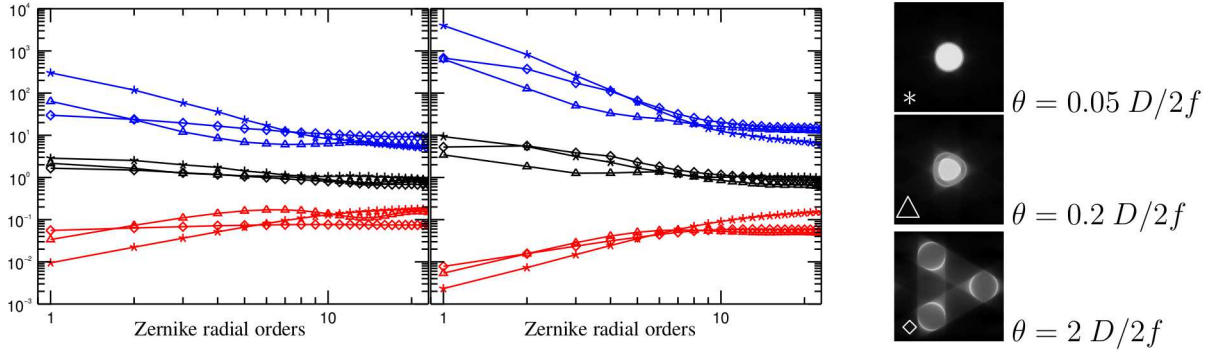}

\caption{Influence of the \textbf{apex angle} on the \textcolor{red}{\textbf{Sensitvity}}, the \textcolor{blue}{\textbf{Linearity range}} and the \textbf{SD factor} with respect to the spatial frequencies in terms of Zernike Radial Orders. Number of faces equals to 3.
Left insert: no modulation. Right insert: circular modulation ($r_m=3 \lambda/D$). \label{2}}
\end{figure}

\begin{figure}[htpb]
\begin{minipage}[c]{0.75\linewidth}
\centering
\includegraphics[width=12cm]{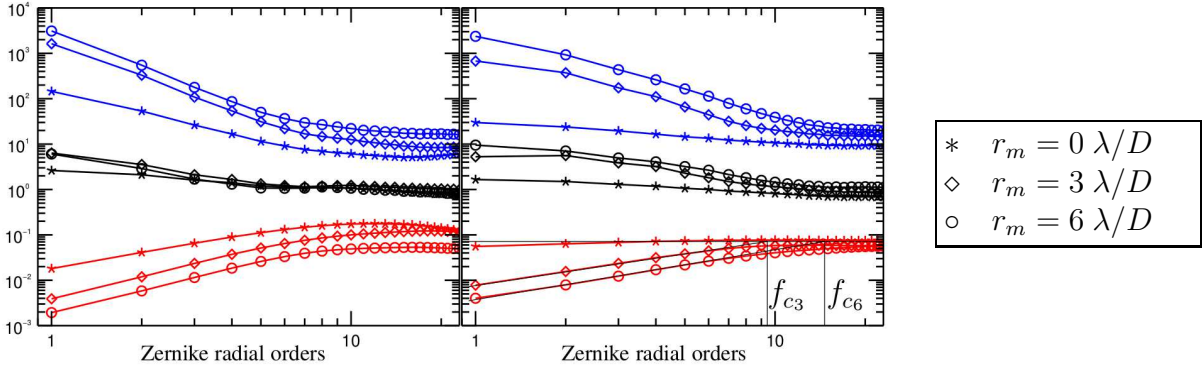}
\end{minipage}~
\fbox{\begin{minipage}[c]{0.2\linewidth}
$*$~~~~$r_m=0~\lambda/D$

$\diamond$~~~~$r_m=3~\lambda/D$

$\circ$~~~~$r_m=6~\lambda/D$
\end{minipage}}
\vspace{0.2cm}

\caption{Influence of the \textbf{modulation radius} on the \textcolor{red}{\textbf{Sensitvity}}, the \textcolor{blue}{\textbf{Linearity range}} and the \textbf{SD factor} with respect to the spatial frequencies in terms of Zernike Radial Orders. Number of faces equals to 3. The modulation is \textbf{circular}. Left insert: small apex angle ($\theta=0.1\frac{D}{2f}$). Right insert: large apex angle ($\theta=2\frac{D}{2f}$). \label{3}}
\end{figure}

\begin{figure}[htpb]
\centering
\includegraphics[width=15cm]{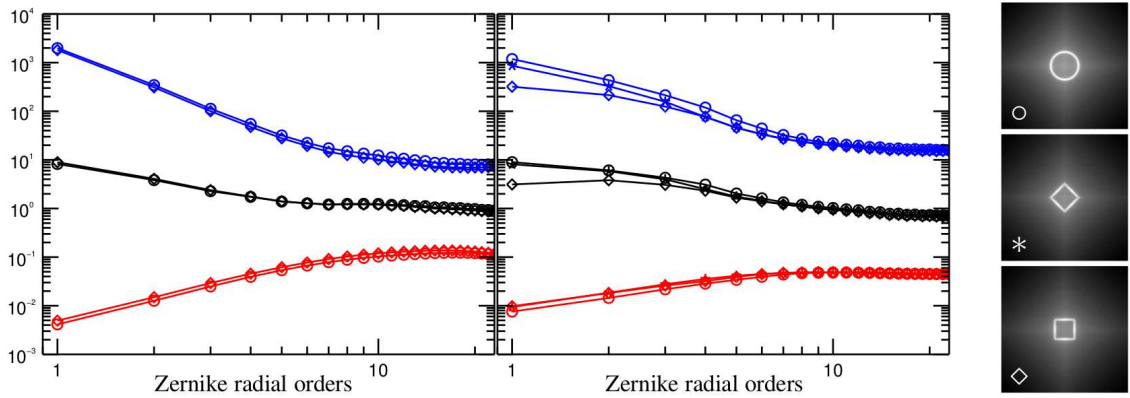}

\caption{Influence of the \textbf{shape of the modulation path} on the \textcolor{red}{\textbf{Sensitvity}}, the \textcolor{blue}{\textbf{Linearity range}} and the \textbf{SD factor} with respect to the spatial frequencies in terms of Zernike Radial Orders. Number of faces equals to 4. Maximum modulation radius equals to $3\lambda/D$ Left insert: small apex angle ($\theta=0.1\frac{D}{2f}$). Right insert: large apex angle ($\theta=2\frac{D}{2f}$). \label{4}}
\end{figure}

\begin{figure}[htbp]
\centering
\includegraphics[width=15cm]{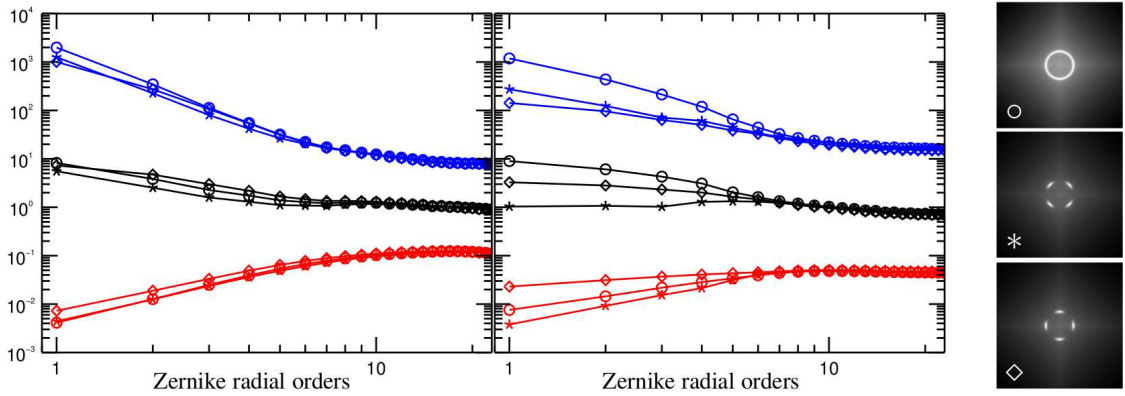}

\caption{Influence of the \textbf{weighting function} on the \textcolor{red}{\textbf{Sensitvity}}, the \textcolor{blue}{\textbf{Linearity range}} and the \textbf{SD factor} with respect to the spatial frequencies in terms of Zernike Radial Orders. Number of faces equals to 4. The modulation is \textbf{circular}. $r_m=3\lambda/D$ Left insert: small apex angle ($\theta=0.1\frac{D}{2f}$). Right insert: large apex angle ($\theta=2\frac{D}{2f}$). \label{5}}
\end{figure}

This section is interested in the influence of the optical and modulation parameters on the \textcolor{red}{\textbf{Sensitvity}}, the \textcolor{blue}{\textbf{SD factor}} and the \textbf{Linearity range} with respect to the spatial frequencies in terms of Zernike Radial Orders. 
Various configurations are shown on figures \ref{1}, \ref{2}, \ref{3}, \ref{4} and \ref{5}.

Figure \ref{1} is interested in the influence of the number of faces. We observe that this optical parameter does not really impact the performance (in terms of $\textcolor{red}{\textbf{s}}$, $\textbf{d}$ and $\textcolor{blue}{\textbf{s.d}}$) of the considered WFSs whatever the apex angle is. Physically, it means that the number of elements in which the spatial frequencies plane is cut does not really matter. Consequently, using a 3-faces pyramid instead of a 4-faces would be particularly convenient since the 3-faces requires less pixels and might be easier to manufacture.

We study on figure \ref{2}, the influence of the apex angle on a 3-faces pyramid with ($r_m=3\lambda/D$) and without modulation. First of all, one can note that as soon as $2f\theta/D>1.5$, the 3 curves do not evolve anymore and this is a general behavior observed for all the configurations tested in this article. In other words, $\theta$ has an influence only when the pupil images overlap. Secondly, as mentioned in Fauvarque et al.\cite{Fauvarque2015} when there is no modulation (left graph of figure \ref{2}) we observe that an optical recombination induced by a small angle provides a better sensitivity in the high spatial frequencies while it decreases for the low frequencies. Moreover, it is possible to choose where the sensitivity is maximum by changing the $\theta$ value. It comes as no surprise that the linearity range has an inverse behavior: it improves for the low frequencies and decreases at the high ones. The curve of the SD factor is more interesting since for small angles, this curve stays above the pyramid with completely separated pupil images for all the spatial frequencies. In terms of trade-off between sensitivity and linearity range, there is thus a real gain to use small angles.   Finally, one can note (right graph of figure \ref{2}) that the modulation has the same influence for a flattened or a classical pyramid before the cut-off spatial frequency: an improvement of the linearity range  frequency associated to a loss of sensitivity. 

We mention here another kind of trade-off which could be particularly relevant in Extreme AO. Indeed, we just observed that the improvement in terms of sensitivity for the high spatial frequencies when the apex angle is small is always associated to a deteriorated sensitivity for the low spatial frequencies. It is not necessarily a problem if the optimization consists in maximizing the Strehl ratio since the cumulative sensitivity is better for flattened pyramids than a classical one. Nevertheless, if the goal is to improve contrast at small angles, such optical configuration may not be relevant since areas close to the center of the point spread function corresponds to low spatial frequencies where the sensitivity is reduced.

Figure \ref{3} is interested in the influence of the modulation radius on a flattened and a classical pyramid. Modulation is here considered as uniform and circular. One can observe on the right graph where the pupil images are completely separated, the classical influence (see Guyon\cite{guyon2005limits} and Verinaud \cite{verinaud2004nature}) of the modulation radius on the sensitivity and the linearity range: an improvement of the linearity range associated to a loss a sensitivity for the low spatial frequencies. Two behaviors are separated by a cut-off frequency  which is growing linearly with the modulation radius: $f_{c_3}$ (resp. $f_{c_6}$) for a modulation radius equals to $3\lambda_D$ (resp. $6\lambda/D$). These two behaviors are a slope sensor for the low frequencies (linear growth of the sensitivity) and a phase sensor (flat sensitivity) for the high frequencies. The linearity range is also improving with the modulation radius. By looking at the SD factor which characterizes the trade-off sensitivity/linearity range, it appears that the gain is particularly pronounced for the low spatial frequencies but tends to be null for the highest ones.  \\
If we consider the case of a flattened pyramid (left graph), one can observe that the modulation radius does not change the spatial frequency where the sensitivity is maximum. Nevertheless, we can note a global improvement of the linearity range associated to a global loss of sensitivity on the whole range of spatial frequencies. Modulation has thus influence for all the spatial frequencies while considering small apex angles.\\ 
To summarize, one can say that effects of modulation and small apex angles add to each other. As a consequence, a WFS with a small apex angle to a high modulation will be much more linear than a classical pyramid with no modulation.

Figure \ref{4} is interested in the influence of the shape of the modulation path. Modulation is considered as uniform and the maximum modulation radius equals to $3\lambda/D$. The curves show that the shape of the path does not really impact the performance whatever the apex angle is.

Figure \ref{5} is finally interested in the influence of the weighting function $w(s)$. By changing this function, it is possible to spend more time on the edges or on the faces. The SD factors of the two graphs of figure \ref{5}  clearly shows that the uniform modulation corresponds to the best trade-off Sensitivity/Linearity range. However, the most sensitive configuration is the one which spends a maximum of time on the edges of the pyramid while the "faces" modulation presents the worst sensitivity. This result is not surprising since the "edges" modulation corresponds to the case which is the closest to the non-modulated Pyramid. 

\section{Conclusions}

The numerous parameters of the sensors of the Pyramid class thus allows to consider the \textbf{optimization} of the element \emph{sensor} in an Adaptive Optics loop. The comparison of the different WFSs of this class, following unified performance criteria, shows that:

-- The \textbf {number of faces} is essentially a geometric and technological parameter. Indeed, this parameter has no influence on the sensitivity and the linearity range. Consequently, its choice has to be done regarding sampling and manufacturing criteria. If the pupil images are completely separated, the 3-faces mask seems the best candidate since it minimizes the number of required pixels and is easy to manufacture. The Cone seems particularly relevant for the small apex angles. It is indeed perfectly chromatic due to the absence of edges and does not require a lot of pixels.

-- The \textbf{tip/tilt modulation stage} provides two main advantages. It allows to retrieve a great part of the photon lost due to diffraction when the pupil images are separated. Moreover it improves drastically the linear range for the low spatial frequencies. Unfortunately a loss of sensitivity is associated to this gain. Simulations show that the shape of the modulation path (squared or circular) has no significant influence. Nevertheless, we observe that the sensitivity was linked to the time spent on the edges whereas the linearity range was correlated to the time spent on faces. Finally we note that modulation inevitably makes the WFS sensible to polychromatism. Without modulation all the sensors of the Pyramid class are rigorously achromatic. 

-- The \textbf{apex angle} turns out to be the most promising parameter of the Pyramid class. Two regimes appear. The first one corresponds to the classical case where the pupils images are completely separated, i.e. $2f\theta/D>1$. In this regime, all the performance criteria are stable regarding to $\theta$. Without modulation it corresponds to the classical \emph{flat} sensitivity and linearity range whatever the number of faces equals to. The second regime corresponds to the flattened Pyramids, i.e. $0<2f\theta/D<0.5$. On this range, the influence of $\theta$ is significant especially concerning the sensitivity and the linearity range. In particular, this parameter allows to choose where the gain in terms of sensitivity is maximal. Moreover, the small angle configurations allows to drastically improve the efficiency of the incoming photon without needing modulation. At the same time, one notes that few pixels are needed to do the wave front sensing.

\noindent The next steps of this study consist in comparing all the WFSs of the Pyramid class on a unique \textbf{optical bench}. Following the results of Akondi et al.\cite{Akondi2013a} the technological device allowing to create all the Fourier masks will be a Spatial Light Modulator. Concerning the forthcoming numerical simulations, we plan to create an AO loop where all the optical parameters of the Pyramid WFS would be, as the deformable mirror, controlled and try to close it depending on the AO contexts. 

\noindent Finally, we mention that this study does not concern the impact of the high-orders spatial frequencies which are not corrected in a AO loop. However, it appears (Korkiakoski et al.\cite{Kork08,Korkk08}) that non-linearity caused by such residual phases has strong impacts on performance. Theoretical investigations will thus be led in this direction by extending our model in order to be able to handle both non null static reference phase and dynamic residuals phase.

\acknowledgments 
The work was partly funded by the European Commission under FP7 Grant Agreement No. 312430 Optical Infrared Coordination
Network for Astronomy and by the French Aerospace Lab (ONERA).
B. Neichel and O. Fauvarque acknowledge the financial support from the French ANR program WASABI to carry out this
work.

\bibliography{library}   

\begin{thebibliography}{10}

\bibitem{Fauvarque16}
O.~Fauvarque, B.~Neichel, J.-F. Fusco, Thierry~Sauvage, and O.~Girault,
  ``{General formalism for Fourier based Wave Front Sensing},'' {\em Optica}
  {\bf 3}(2)  (2016).

\bibitem{zer1934}
F.~Zernike, ``{Diffraction theory of the knife-edge test and its improved form,
  the phase-contrast method.},'' {\em Royal Astronomical Society} {\bf 94},
  377--383  (1934).

\bibitem{Rag96}
R.~Ragazzoni, ``{Pupil plane wavefront sensing with an oscillating prism},''
  {\em Journal of Modern Optics} {\bf 43}, 289--293  (1996).

\bibitem{Akondi2014}
V.~Akondi, S.~Castillo, and B.~Vohnsen, ``{Multi-faceted digital pyramid
  wavefront sensor},'' {\em Optics Communications} {\bf 323}, 77--86  (2014).

\bibitem{Vohnsen2011}
B.~Vohnsen, S.~Castillo, and D.~Rativa, ``{Wavefront sensing with an
  axicon.},'' {\em Optics letters} {\bf 36}(6), 846--848  (2011).

\bibitem{Fauvarque2015}
O.~Fauvarque, B.~Neichel, T.~Fusco, and J.-F. Sauvage, ``{Variation around a
  pyramid theme : optical recombination and optimal use of photons},'' {\em
  Optics Letters} {\bf 40}(15), 3528--3531  (2015).

\bibitem{LBT}
S.~{Esposito}, A.~{Riccardi}, E.~{Pinna}, A.~T. {Puglisi},
  F.~{Quir{\'o}s-Pacheco}, C.~{Arcidiacono}, M.~{Xompero}, R.~{Briguglio},
  L.~{Busoni}, L.~{Fini}, J.~{Argomedo}, A.~{Gherardi}, G.~{Agapito},
  G.~{Brusa}, D.~L. {Miller}, J.~C. {Guerra Ramon}, K.~{Boutsia}, and
  P.~{Stefanini}, ``{Natural guide star adaptive optics systems at LBT: FLAO
  commissioning and science operations status},'' in {\em Adaptive Optics
  Systems III},  {\em Proc. SPIE} {\bf 8447}, 84470U  (2012).

\bibitem{MagAO}
L.~M. {Close}, J.~R. {Males}, D.~A. {Kopon}, V.~{Gasho}, K.~B. {Follette},
  P.~{Hinz}, K.~{Morzinski}, A.~{Uomoto}, T.~{Hare}, A.~{Riccardi},
  S.~{Esposito}, A.~{Puglisi}, E.~{Pinna}, L.~{Busoni}, C.~{Arcidiacono},
  M.~{Xompero}, R.~{Briguglio}, F.~{Quiros-Pacheco}, and J.~{Argomedo},
  ``{First closed-loop visible AO test results for the advanced adaptive
  secondary AO system for the Magellan Telescope: MagAO's performance and
  status},'' in {\em Adaptive Optics Systems III},  {\em Proc. SPIE} {\bf
  8447}, 84470X  (2012).

\bibitem{Subaru}
N.~{Jovanovic}, O.~{Guyon}, F.~{Martinache}, C.~{Clergeon}, G.~{Singh},
  T.~{Kudo}, K.~{Newman}, J.~{Kuhn}, E.~{Serabyn}, B.~{Norris}, P.~{Tuthill},
  P.~{Stewart}, E.~{Huby}, G.~{Perrin}, S.~{Lacour}, S.~{Vievard},
  N.~{Murakami}, O.~{Fumika}, Y.~{Minowa}, Y.~{Hayano}, J.~{White}, O.~{Lai},
  F.~{Marchis}, G.~{Duchene}, T.~{Kotani}, and J.~{Woillez}, ``{Development and
  recent results from the Subaru coronagraphic extreme adaptive optics
  system},'' in {\em Ground-based and Airborne Instrumentation for Astronomy
  V},  {\em Proc. SPIE} {\bf 9147}, 91471Q  (2014).

\bibitem{ragazzoni1999sensitivity}
R.~Ragazzoni and J.~Farinato, ``Sensitivity of a pyramidic wave front sensor in
  closed loop adaptive optics,'' {\em Astronomy and Astrophysics} {\bf 350},
  L23--L26  (1999).

\bibitem{Esposito1999}
S.~Esposito and A.~Riccardi, ``Pyramid wavefront sensor behavior in partial
  correction adaptive optic systems,'' {\em Astronomy and Astrophysics} {\bf
  369}(2), L9--L12  (2001).

\bibitem{verinaud2004nature}
C.~V{\'e}rinaud, ``On the nature of the measurements provided by a pyramid
  wave-front sensor,'' {\em Optics Communications} {\bf 233}(1), 27--38
  (2004).

\bibitem{guyon2005limits}
O.~Guyon, ``Limits of adaptive optics for high-contrast imaging,'' {\em The
  Astrophysical Journal} {\bf 629}(1), 592  (2005).

\bibitem{Akondi2013a}
V.~Akondi, S.~Castillo, and B.~Vohnsen, ``{Digital pyramid wavefront sensor
  with tunable modulation.},'' {\em Optics express} {\bf 21}(15), 18261--18272
  (2013).

\bibitem{Kork08}
V.~Korkiakoski, C.~V\'{e}rinaud, and M.~L. Louarn, ``Improving the performance
  of a pyramid wavefront sensor with modal sensitivity compensation,'' {\em
  Appl. Opt.} {\bf 47}, 79--87  (2008).

\bibitem{Korkk08}
V.~{Korkiakoski}, C.~{V{\'e}rinaud}, and M.~{Le Louarn}, ``{Applying
  sensitivity compensation for pyramid wavefront sensor in different
  conditions},'' in {\em Adaptive Optics Systems},  {\em Proc. SPIE} {\bf
  7015}, 701554  (2008).

\end{thebibliography}
\bibliographystyle{spiejour}   

\listoffigures

\end{spacing}
\end{document}